\documentstyle[preprint,prl,aps]{revtex}          
\begin{document}

\author{R. Cowsik  \\ 
{\it Indian Institute of Astrophysics,}\\
{\it Sarjapur Road, \ Bangalore - 560034, India}\\{\it and}\\{\it Tata Institute of
Fundamental Research, }\\{\it Homi Bhabha Road, Mumbai - 400005, India}}

\title{Radio Ranging Techniques to Test Relativistic Gravitation}
\maketitle
\begin{abstract}
\baselineskip=18pt It is suggested that modern techniques of radio ranging
when applied to study the motion of the Moon, can improve
the accuracy of tests of relativistic gravitation obtained with  currently
operating laser ranging techniques.  Other auxiliary information relevant
to the Solar system would also emerge from such a study.

\smallskip\

\end{abstract}

\baselineskip=20pt

\vskip .5cm

Eversince Newton used precisely known orbit of the Moon to substantiate
his theory of Universal gravitation, the study of its orbit has provided
continual impetus in the formulation and testing of theories of gravity.
Most remarkable of the modern developments in this context is the
discovery by Nordvett that the passive gravitational mass of an object may
not be the same as the inertial mass in the most general metric theories
of gravity[1-3]   Thus there exists the
possibility, referred to as the Nordvett effect, that bodies with
gravitational self-energy may in principle violate the Weak Equivalence
Principle or the Universality of free-fall in a given gravitational field.
 Motivated in part by the desire to test the Nordvett-effect laser
retroreflectors were placed on the Moon in 1969 and later again in 1971 by
the Apollo-astronauts[4].  In 1973 the French-Russian collaboration added
some more reflectors.  These reflectors are used regularly eversince to obtain
precise distances to the Moon by laser ranging.  A laser pulse of width of
$\sim$ 200 psec. comprising of $\sim$  $10^{18}$ photons is beamed towards
the Moon by an Earth-based optical telescope which also attempts to record
the reflected photons after the two-way travel $\Delta$t time of about 2.9
seconds.  The distance to the Moon is given by C$\Delta$t/2, and the
accuracy of this measurement (based on an ensemble of reflected photons)
has been steadily improving from about 25 cm in the 1970's to about 3 cm
at present[5].

\vskip .5cm

Keeping in mind that under laboratory conditions, one can use lasers to
determine spacings to a millionth of the wavelength of light, we may
inquire as to why the accuracies of Lunar Laser Ranging(LLR) are limited
to the values stated here.  The primary reason for this is the extreme
weakness of the signals received at the Earth station after reflection of
the laser pulse at the Moon.  In its outward journey the intensity of the
photons decreases inversely as the square of the distance because of
finite beam divergence $\geq 10^{-5}$ radians; consequently only a tiny
fraction of these photons are intercepted by the retroreflectors.  The
photons reflected from the retroreflectors diverge even more than the
incoming photons and suffer a decrease in intensity by another factor of
r$^{-2}$.  Thus the intensity of the received photons drops as r$^{-4}$ so
that only about 1 photon in about 3-4 seconds is recorded after the
two-way travel.  In the radio methods which we describe here the loss of
intensity is confined to only one-way propagation so that the intensity
drops only as $\sim \thinspace r^{-2}$.  These methods allow considerably
greater accuracy and continuous coverage in performing the ranging
observations.

\vskip .5cm

Radio Ranging Techniques (RRT) are able to ameliorate this problem in
several ways, primarily by not having to reflect the received signal from
the Moon back to the Earth.  Some of these methods were developed for
establishing the 
Global Positioning System (GPS)[6].  Two carriers are broadcast from each of
several transmitting stations stamped with a code at regular intervals
giving the station location and accurate time, based on stable atomic
clock.  The receiver has a clock of modest stability which generates an
identical code and cross-correlates this
with the received signal to determine the time at reception and the phase
of the carrier signal.  After correcting for the special relativistic
Doppler shift in frequency due to the relative motion of the satellites
and the receiver station and the general
relativistic change in frequency due to the difference in the
gravitational potentials, one obtains the pseundorange and
phase of the signal for each station which are still to be corrected for
dispersion during propagation, clock errors and so on. The procedure used for
deconvolving the received data to obtain the actual vector coordinates of
the GPS receivers are described in detail in the book by A.Leick and
in the IEEE Proceedings[6].  It suffices to state here that even with the
existing set of GPS satellites which broadcast at $\sim$ 1-1.5 GHz
($\lambda$=20-30 cm) a differential positional accuracy of $\sim$ 1-2 mm may be
obtained with in several hours of observation using commercially available
top of the line GPS receivers.  One should emphasize that this accuracy is
for a differential measurement with respect to some fixed station albeit a few
thousand kilometers away.  The contributors to the error budget in such a
measurement are many, the uncorrected part of propagation delays due to
the finite refractive indices of the ionosphere and the troposphere,
multiple path effects, uncertainties in the orbital location of the
satellites, finite signal to noise ratio of the reception and so on.  At
the same time one should recognize that it is the 3-dimensional position of the
receiver which is measured.  When GPS-type receivers are placed on the
Moon and the same set of satellites are used for Lunar Radio Ranging (LLR)
then because of the larger distances involved the radio intensities would
considerably be weaker and correspondingly the signal to noise ratio will
drop below the value achieved by the Earth-bound receivers. This drop in
S/N ratio is by about a factor of 400 so that one has to integrate for
several days to achieve the same level of accuracies in position
determination.  Notice however the advantage of radio ranging over laser
ranging in that the GPS receiver placed on the Moon determines its own
position and velocity by merely receiving the signals from the several
transmitters and there is no need to reflect the incoming signal as in
the case of laser ranging, with concommittant decrease in the signal
strength and the achieved accuracies in position determination.  All that
the radio instrument on the Moon does is to send by telemetry at regular
intervals the position and velocity determined during each cycle of
integration.  This means that in radio ranging the signals drop only as
$r^{-2}$ and not as severely as $r^{-4}$ as for laser ranging. Further it
is possible to deploy GPS-type receivers with dish 
antennas of about 1 m in diameter and thus improve the signal to
noise ratio of the reception and consequently the measurement accuracy
considerably.  

\vskip .5cm
It is perhaps easier to design a dedicated LRR with several antennas of
moderate size deployed at suitably chosen stations on the Earth tuned at
centimetre wavelengths which beam their signals at the receivers on the
Moon.  The cm wavelengths would have the advantage of relatively small
ionospheric contribution and negligible  trophospheric attenuation due to
waver vapour.  In such a system, it would be beneficial to have a
transponder which would suitably shift the signal to a reduced downlink
frequency and amplify the signal before re-transmitting to the Earth.  By
comparing the phases of the received signals with those of the transmitted
signals, it is possible to determine the velocity of the Moon with respect
to the Earth station.  By using multiple frequencies and phase-comparing
difference frequencies, the system can be made immune to random phase
jitter at the translation oscillator and clock errors.  By employing such
a technique on a geostationary satellite the line of sight velocity of the
satellite has been measured recently to an accurary of 0.1 cm/s with a few
seconds of integration (N.V.G. Sarma and C.R.Subrahmanya, Private
Communication). Thus from these signals which reach
the Moon with negligible dispersion in the ionosphere (because the
dispersion drops as $\nu^{-2}$) each receiver will be able to determine
its vector position and velocity with considerably greater accuracy than
what has been possible with LLR.
The point that is being made here is not that there is any fundamental
difference between the use of radio-waves in contrast with optical photons
of the lasers; it is just that in its present form the radio technique
incorporates the well established technology of ``intelligent'' analysis
of the signal received on the Moon by the receiver.  In principle, a
similar technique could be achieved with the laser signals also.

\vskip .5cm

Despite the r$^{-4}$ drop in the laser signal, which limits the LLR
capabilities the extraordinarily long stretch of the data string extending
over more than two decades has provided the most stringent constraints as
yet on the Universality of free-fall and the independence of this
acceleration on the gravitational self-energy content of the body.  To
review these note that the LLR observations yield[5,7,8]

\begin{equation}
\label {1} \frac {1}{2} \left| \frac {a_{\text{Earth}} - a_{\text
{Moon}} } {a_{\text{Earth}} + a_{\text {Moon}} } \right| \leq \thinspace
5\times 10^{-13}
\end{equation}

\noindent where `$a_i$' represent the accelerations suffered by the
bodies in the gravitational field of the Sun.  Since the gravitational
self energy of the Earth is a greater fraction of its restmass compared to
that of the Moon, with

\begin{equation}
\label {2} \left| \left( \frac {U}{mc^2} \right)_{\text {Earth}} - \left(
\frac {U}{mc^2}\right)_{\text {Moon}} \right| = 4.5 \times 10^{-10}
\end{equation}

\noindent the constraint on $\eta$ representing any dependence on the
gravitational accelerations on self energy turns out to be

\begin{equation}
\label{3} \eta \thinspace \leq \thinspace 10^{-3}
\end{equation}

This would represent a test of the Nordvett effect at the $10^{-3}$ level,
only when we can rule out a fortuitous cancellation of the effect by a
corresponding violation of the weak equivalence principle induced by the
differences in the composition of the Earth and the Moon.  Since the
laboratory experiments[9,10] rule out such composition dependent effects only 
at the $5 \times 10^{-12}$ level the LLR data, without any extra assumptions
yield the strict bound

\begin{equation}
\label {4} \eta = (4 \beta - \gamma - 3) \leq 10^{-2}
\end{equation}

\noindent
where $\beta$ and $\gamma$ are the standard PPN-parameters[3].

Considering the orbital angular momentum vector of the Moon to represent a
gyroscope the LLR data also confirms the geodetic precession of this
gyroscope in the field of the Sun to within 0.7 \% of the theoretically
estimated value of 19.2 milliarcsecond per year.

\vskip .5cm

Finally LLR imposes the best limits on any secular decrease in the value
of the gravitational constant, say according to Dirac's large number
hypothesis.  Noting that such a decrease will make the distance to the
Moon and its period of orbital motion increase with time, the variation is
bounded by[5,7,8] 

\begin{equation}
\label {5} \stackrel {.}{G}/\stackrel{}{G} = (1 \pm 3) \times 10^{-12} /y
\thinspace < \thinspace  7 \times 10^{-12} / y.
\end{equation}

The exact level of improvement that we may expect with radio ranging
techniques in improving these bounds on the
deviations from General Relativity would clearly depend on the accuracy
with which we can predict the ``Standard Orbit'' of the Moon under the
assumption of standard gravitation.  Uncertainties such as the tidal torque on
the Moon which depends on its internal structure, the perturbation by
planets and other bodies etc. may finally impose limits on the accuracy
with which one can
derive the PPN parameters.  However, it is not unreasonable to hope that
the study carried out over an extended period may not only improve
considerably our knowledge of the parameters \mbox {$\beta, \thinspace
\gamma,\thinspace \stackrel {.}{G}/ \stackrel{}{G},$} etc. but may also lead
to a better understanding of the perturbing torques and
forces in the solar system.  In closing, we wish to remark that there are
several experimental efforts underway which will attempt to test the
weak equivalence principle in the laboratory and in space[11,12] so that
the improved LRR results can be explicitly searched for the violation SEP
and the Nordvett effect. In closing we would like to stress the importance
of studies in gravitation keeping in mind the possible deep
interconnections this field bears with the other fundamental fields of
physics[13].

\newpage\

\noindent {\bf Acknowledgment}

\vskip .5cm

It is indeed a great pleasure to thank my colleagues Dr V.Radhakrishnan,
Dr R.Nityananda, Dr C.R.Subrahmanya Dr N.Krishnan and Dr C.S.Unnikrishnan who gave their
time generously for the discussion of the ideas presented in this paper.

\vskip 2cm

\vfill\eject

\end{document}